\documentclass[12pt,epsf,fleqn]{article}
\usepackage{epsfig}
\usepackage[dvips]{color}
\begin{document}
\title{Possibility of spontaneous CP violation in Higgs physics beyond the minimal supersymmetric standard model}
\author{S. W. Ham$^{(1)}$, Seong-A Shim$^{(2)}$, and S. K. Oh$^{(3)}$
\\
\\
{\it (1) Department of Physics, Korea University, Seoul 136-701} \\
{\it (2) Department of Mathematics, Sungshin Women's University} \\
{\it  Seoul 136-742} \\
{\it (3) Department of Physics, Konkuk University, Seoul 143-701} \\
\\
\\
}
\date{}
\maketitle
\begin{abstract}
The Dine-Seiberg-Thomas model (DSTM) is the simplest version of the new physics beyond the minimal
supersymmetric standard model (MSSM), in the sense that its Higgs sector has just two dimension-five operators,
which are obtained from the power series of the energy scale for the new physics in the effective action analysis.
We study the possibility of spontaneous CP violation in the Higgs sector of the DSTM, which consists of two Higgs doublets.
We find that the CP violation may be triggered spontaneously by a complex phase, obtained as the relative
phase between the vacuum expectation values of the two Higgs doublets.
At the tree level, for a reasonably established parameter region,
the masses of the three neutral Higgs bosons and their corresponding coupling coefficients to a pair of $Z$ bosons in the DSTM
are calculated such that the results are inconsistent with the experimental constraint by the LEP data.
Thus, the LEP2 data exclude the possibility of spontaneous CP violation in the DSTM at the tree level.
On the other hand, we find that, for a wide area in the parameter region, the CP symmetry may be
broken spontaneously in the Higgs sector of the DSTM at the one-loop level,
where top quark and scalar top quark loops are taken into account.
The upper bound on the radiatively corrected mass of the lightest neutral Higgs boson of the DSTM is about 87 GeV,
in the spontaneous CP violation scenario.
We confirm that the LEP data does not exclude this numerical result.
\end{abstract}
\vfil\eject

\section{Introduction}

Recently, Dine, Seiberg, and Thomas have investigated the effects of new physics beyond the
minimal supersymmetric model (MSSM) within the framework of effective field theory analysis [1].
They have shown that, in general, if the new physics beyond the MSSM lies at an energy scale $M$,
the corrections to the MSSM may be described in terms of higher-dimensional operators.
These higher-dimensional operators emerge from a power series of $1/M$ in the low-energy effective Lagrangian density.

Even though the higher dimensional operators are suppressed by the power of the new physics scale $1/M$,
the leading order effects of these operators on the physical observables may be phenomenologically
comparable to the one-loop effects of some theories beyond the MSSM.
Thus, it is worthwhile studying the implications of these higher dimensional operators in the Higgs phenomenology.
Dine, Seiberg, and Thomas show that the effective dimension of the operators are five or more.
The Higgs sector of the simplest version has just two dimension-five operators with the MSSM particle content,
at the energy scale below $M$.
We may call it hereafter as the Dine-Seiberg-Thomas model (DSTM).

Actually, some interesting effects are observed in the MSSM electroweak baryogenesis and the MSSM fine tuning,
when the MSSM is incorporated with higher dimensional operators.
In the MSSM with higher-dimensional operators, it has been shown that the strength of the electroweak phase transition
might be strongly first order without requiring that the scalar top quark is lighter than the top quark [2].
Also, the burden of fine tuning in the MSSM might be reduced significantly by the effects of the higher dimensional operators [3].
Meanwhile, the dimension-five operators in the MSSM push up the mass of the lightest scalar Higgs boson
up to about 105 GeV from the $Z$ boson mass, which is its tree-level upper bound [4].

It is known that the Next-to-MSSM (NMSSM), which is the most popular one among the
supersymmetric models beyond the MSSM, may solve difficulties of the light stop scenario
for electroweak baryogenesis as well as the fine-tuning problem, for a wide parameter space [5].
The scale of the NMSSM may be a few TeV, near the low energy SUSY breaking scale,
whereas the scale of a new physics might emerge above the TeV scale.
For example, the scale of some gauge-mediated supersymmetric scenarios require $10 - 1000$ TeV [6].
The effective field analysis may be useful in the phenomenological point of view, since
it is valid for a wide range of energy scale from the SUSY breaking scale to the scale of new physics.

We have been interested in the possibilities of CP violation in the MSSM and beyond [7],
since we consider the violation of CP symmetry as one of the important subject for the phenomenology of
low-energy supersymmetric models [8].
In principle, CP violation is induced by the mixing between the scalar and pseudoscalar Higgs bosons
for any model that has at least two Higgs doublets [9].
Supersymmetric standard models, including the MSSM and the DSTM, share this property.

It has been observed that the MSSM has some difficulties in realizing CP violations, although
the complex phases in $\mu$ and the soft SUSY breaking parameters are the possible sources of CP violation.
Explicit CP violation, arising directly from the complex phases in these parameters, is viable in the MSSM
at the one-loop level due to the radiative CP mixing among the scalar and pseudoscalar Higgs bosons [10,11].

Spontaneous CP violation at the tree-level is impossible in the MSSM, since the complex phases in the
vacuum expectation values of the two Higgs doublets may always be eliminated by a global phase rotation [12].
At the one-loop level, the complex phases in the vacuum expectation values of two the Higgs doublets do not cancel
and thus may trigger the spontaneous CP violation in the MSSM.
However, one of the scalar Higgs bosons in the MSSM turns out to be very light,
which is excluded by the LEP data [13].

Since the DSTM has two Higgs doublets, it also has the possibilities of CP violation.
In this article, we study whether the DSTM may accommodate CP violation in its Higgs sector.
We find that the CP violation may occur spontaneously in the Higgs sector of the DSTM at the one-loop level,
without contradicting the negative results of the light Higgs search at LEP2.
The radiative corrections to the tree-level Higgs sector of the DSTM are calculated by taking into account
the top and scalar top quark loop contributions.

\section{Higgs Sector}

Let us study the Higgs sector of the DSTM, which consists of the two Higgs doublets,
\begin{eqnarray}
& & H_d = \left(
\begin{array}{c}
H_d^0 \\
H_d^-
\end{array}
\right)
= \left (
\begin{array}{c}
v_d e^{i \varphi_d} + \phi_d + i \psi_d \\
H_d^-
\end{array}
  \right )
\ , \cr
& & H_u= \left(
\begin{array}{c}
H_u^+ \\
H_u^0
\end{array}
\right)
= \left (
\begin{array}{c}
H_u^+ \\
v_u e^{i \varphi_u} + \phi_u  + i \psi_u
\end{array}
\right )   \ ,
\end{eqnarray}
where $H_d^0$ and $H_u^0$ are the neutral Higgs fields, $H_d^-$ and $H_u^+$ are the charged Higgs fields,
$v_d e^{i \varphi_d}$ and $v_u e^{i \varphi_u} $ are complex vacuum expectation values of the Higgs doublets,
Note that, if CP is conserved in the Higgs sector, the real and the imaginary components of the neutral Higgs fields
would have definite CP parity, namely, $\phi_d$ and $\phi_u$ would be the scalar fields
while $\psi_d$ and $\psi_u$ would be the pseudoscalar fields.
However, since $\varphi_d$ or $\varphi_u$ are generally not zero, giving rise to the possibility of spontaneous CP violation,
the real and the imaginary components of the neutral Higgs fields may mix and thus may not have definite CP parity.

The DSTM has two Higgs doublets like the MSSM, but its Higgs structure is different from the MSSM.
At the tree-level, the general Higgs potential of the DSTM in terms of $H_u$ and $H_d$ is given as
\begin{eqnarray}
V_0 & = & m_u^2 H_u^{\dagger} H_u + m_d^2 H_d^{\dagger} H_d - (m_{ud}^2 H_u H_d + {\rm H.c.} )
 + {\lambda_1 \over 2} (H_u^{\dagger} H_u)^2 + {\lambda_2 \over 2} (H_d^{\dagger} H_d)^2       \cr
&  &\mbox{}
+ \lambda_3 (H_u^{\dagger} H_u) (H_d^{\dagger} H_d) + \lambda_4 (H_u^{\dagger} H_u) (H_d^{\dagger} H_d)       \cr
&  &\mbox{}
+ \left [{\lambda_5 \over 2} (H_u H_d)^2
+ \left \{ \lambda_6 (H_u^{\dagger} H_u) + \lambda_7 (H_d^{\dagger} H_d) \right \} H_u H_d + {\rm H.c.} \right ]  \  ,
\end{eqnarray}
where $m_d^2 \equiv m_{H_d}^2 + |\mu|^2$, $m_u^2 \equiv m_{H_u}^2 + |\mu|^2$,
$m_{ud}^2 \equiv - \mu B$, and $\lambda_i$ ($i$ = 1-7) are the quartic couplings.
They are defined as
\begin{eqnarray}
& & \lambda_1 = \lambda_2 = {1 \over 4} ({g'}^2 + g^2), \quad \lambda_3 = {1 \over 4} (g^2 - {g'}^2) \ , \cr
& & \lambda_4 = - {1 \over 2} g^2 \ , \quad \lambda_5 = 2 \epsilon_2 \ , \quad \lambda_6 = \lambda_7 = 2 \epsilon_1   \  ,
\end{eqnarray}
where $g'$ and $g$ are respectively the gauge coupling coefficients of $U(1)_Y$ and $SU(2)_L$,
and $\epsilon_1$ and $\epsilon_2$ are the coupling coefficients representing the interactions of two dimension-five operators.
Note that $m_d$ and $m_u$ may be eliminated by the two minimum conditions that define the vacuum
with respect to $\phi_d$ and $\phi_u$.

In the MSSM, the quartic couplings are given in terms of the electroweak gauge coupling coefficients alone.
Thus, the upper bound on the tree-level mass of the lightest scalar Higgs boson is determined by $Z$ boson mass.
Consequently, we need large radiative corrections to push up the tree-level mass of the lightest scalar Higgs boson.
On the other hand, the DSTM has two dimension-five operators, which provide higher-dimensional interactions
besides the gauge couplings.

After spontaneous symmetry breaking, $V_0$ develops two complex vacuum expectation values,
$v_d e^{i \varphi_d}$ and $v_u e^{i \varphi_u} $,
physically, only one non-trivial CP phase $\varphi = \varphi_d + \varphi_u$ remains.

\subsection{Tree level}

In spontaneous CP violation scenario, the vacuum must be stable with respect to $\varphi$.
The minimum equation for $\varphi$ at the tree level yields an expression for the mass parameter $m_{ud}$ as
\begin{equation}
m_{ud}^2 = 2 v^2 \left (\epsilon_1 + \epsilon_2 \sin 2 \beta \cos \varphi \right )    \   ,
\end{equation}
where $v  = \sqrt{v_d^2 + v_u^2} = 175$ GeV and $\tan \beta = v_u/v_d$.
After eliminating the mass parameters $m_d$, $m_u$, and $m_{ud}$ in the tree-level Higgs potential,
one may obtain the $3 \times 3$ symmetric mass matrix for the neutral Higgs bosons as
\begin{equation}
     M^0 =
    \left ( \begin{array}{cccc}
    M_{11}^0 & M_{12}^0 & M_{13}^0  \cr
    M_{12}^0 & M_{22}^0 & M_{23}^0  \cr
    M_{13}^0 & M_{23}^0 & M_{33}^0  \cr
        \end{array}
    \right ) \ ,
\end{equation}
where
\begin{eqnarray}
M_{11}^0 & = & m_Z^2 \cos^2 \beta + 4 \epsilon_1 v^2 \sin 2 \beta \cos \varphi
+ 2 \epsilon_2 v^2 \sin 2 \beta \tan \beta \cos^2 \varphi   \ , \cr
M_{22}^0 & = & m_Z^2 \sin^2 \beta + 4 \epsilon_1 v^2 \sin 2 \beta \cos \varphi
+ 2 \epsilon_2 v^2 \sin 2 \beta \cot \beta \cos^2 \varphi   \ , \cr
M_{33}^0 & = & 4 \epsilon_2 v^2 \sin^2 \varphi  \  ,   \cr
M_{12}^0 & = &\mbox{} - m_Z^2 \cos \beta \sin \beta + 4 \epsilon_1 v^2 \cos \varphi
- 2 \epsilon_2 v^2 \sin 2 \beta \sin^2 \varphi   \ , \cr
M_{13}^0 & = &\mbox{} - 4 \epsilon_1 v^2 \cos \beta \sin \varphi
- 2 \epsilon_2 v^2 \sin \beta \sin^2 \varphi   \ , \cr
M_{23}^0 & = &\mbox{} - 4 \epsilon_1 v^2 \sin \beta \sin \varphi
- 2 \epsilon_2 v^2 \cos \beta \sin^2 \varphi   \ ,
\end{eqnarray}
with $m_Z^2 = ({g'}^2+g^2)/2$.
The eigenstates of this mass matrix are the three neutral Higgs bosons $h_i$ ($i=1,2,3$)
and the corresponding eigenvalues are respectively the squared masses $m^2_{h_i}$ ($i=1,2,3$).
The masses are sorted such that $m_{h_1} < m_{h_2} < m_{h_3}$.
The mass eigenstates $h_i$ ($i=1,2,3$) do not have definite CP parity in spontaneous CP violation scenario.

Note that the matrix elements $M_{i3}$ $(i=1,2)$ are responsible for the mixing between the scalar and
pseudoscalar Higgs fields.
It is easy to see that these matrix elements would vanish if $\varphi= 0$.
Thus, the non-zero $\varphi$ triggers spontaneous CP violation.

In order to calculate the masses of the neutral Higgs bosons, we need concrete and plausible numbers for the relevant parameters.
We set up a parameter region, defined as $|\varphi| < \pi/2$, $0 < \epsilon_1 < 0.05$, $0 < \epsilon_2 < 0.05$,
and $2 < \tan \beta < 30$.
Then, we scan $5 \times 10^6$ points in this parameter region, by using the Monte Carlo method.
For each point, we calculate $m_{h_i}$ and $g_{ZZh_i}^2$ ($i = 1,2,3$), where $g_{ZZh_i}^2$ ($i = 1,2,3$) is
the normalized Higgs coupling coefficient to a pair of $Z$ bosons, normalized by the corresponding SM Higgs coupling coefficient.

\setcounter{figure}{0}
\def\figurename{}{}%
\renewcommand\thefigure{FIG. 1}
\begin{figure}[t]
\begin{center}
\includegraphics[scale=0.5]{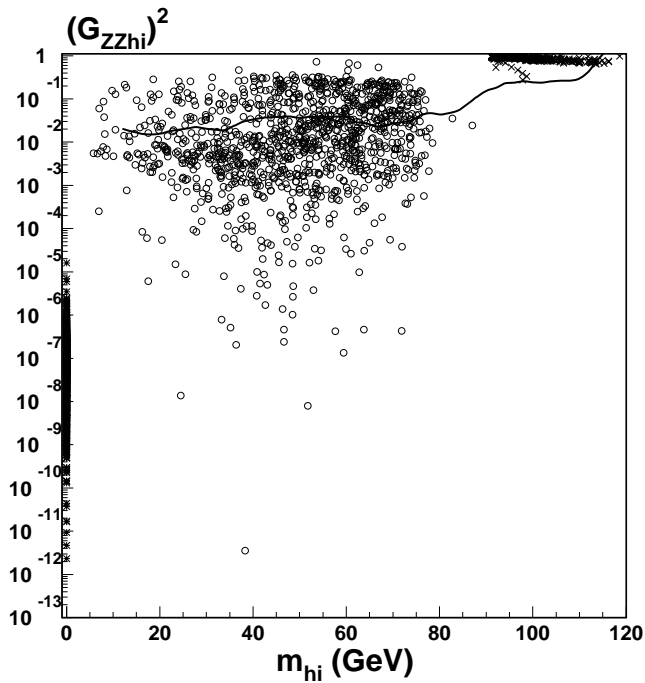}
\caption[plot]{The distribution of $(m_{h_1} , g_{ZZh_1}^2)$ (stars),
$(m_{h_2} , g_{ZZh_2}^2)$ (circles), and $(m_{h_3} , g_{ZZh_3}^2)$ (crosses),
for each of 1145 points in the parameter region, defined as
$|\varphi| < \pi/2$, $0 < \epsilon_1 < 0.05$, $0 < \epsilon_2 < 0.05$,
and $2 < \tan \beta < 30$.
The solid curve is the model-independent upper bound on $g_{ZZH}^2$,
the square of the coupling coefficient of a Higgs boson to a pair of $Z$ bosons,
obtained from the LEP experiments. }
\end{center}
\end{figure}

The result is shown in Fig. 1, where we plot $(m_{h_1} , g_{ZZh_1}^2)$, $(m_{h_2} , g_{ZZh_2}^2)$,
and $(m_{h_3} , g_{ZZh_3}^2)$, for each point in the parameter region.
They are marked respectively by stars, circles, and crosses.
One may notice that the number of points in Fig 1 is
far smaller than $5 \times 10^6$.
This is because some of the $5 \times 10^6$ sets of parameter values randomly
selected in the parameter space yield unphysical results, such as negative masses.
The result is shown in Fig. 1, where a distribution of 1145 points is displayed
in the $(m_{h_i}, g_{ZZh_i}^2)$-plane.
Most of the stars, $(m_{h_1} , g_{ZZh_1}^2)$, are located near the left wall of the figure.
Thus, the lightest neutral Higgs boson is nearly massless: $0 <  m_{h_1} {\rm ~(GeV)~} \le 1$.
On the other hand, most of the crosses, $(m_{h_3} , g_{ZZh_3}^2)$, are gathered to the upper right corner of the
figure such that $91.1 \le m_{h_3} {\rm ~(GeV)~} \le 118.6$.
The crowd of circles for $h_2$ are rather distributed evenly over wide areas in the figure.
Thus, we have $5.8 \le m_{h_2} {\rm ~(GeV)~} \le 87.0$.

One might suspect that the calculated masses of the neutral Higgs bosons are too small to be accepted by the current LEP data.
However, the LEP data should be reinterpreted by taking into account
the coupling coefficient of the SM Higgs boson to a pair of $Z$ bosons.
In other words, a light Higgs boson might be acceptable by the LEP data,
provided that its coupling coefficient to a pair of $Z$ bosons is much smaller than the corresponding SM coupling coefficient.
Thus, it is important to check the size of $g_{ZZh_i}^2$ and $m_{h_i}$, at the same time.

In Fig. 1, we also show a solid curve, as a function of the Higgs boson mass.
It is $(g^{\rm max}_{ZZH})^2$, the model-independent upper bound on the Higgs
coupling coefficient to a pair of $Z$ bosons, obtained from the LEP data [13].
Basically, points below the curve may be accepted by the LEP data, whereas points above the curve may not be accepted.
From Fig. 1, it is clearly seen that most of crosses, $(m_{h_3} , g_{ZZh_3}^2)$, are located above the solid curve,
and thus they should be rejected by the LEP data.
This implies that most points in the parameter region are not consistent with the LEP data,
with respect to spontaneous CP violation at the tree level.

There are a few crosses of $(m_{h_3} , g_{ZZh_3}^2)$ below the solid curve, especially for $m_{h_3} < 115$ GeV.
This may indicate that a few points in the parameter region is allowed by the LEP data.
However, this is not the case.
We find that the points in the parameter region that produce these crosses produce $(m_{h_2} , g_{ZZh_2}^2)$ above the solid curve.
Therefore, we find that the randomly chosen 1145 points in the parameter region are all excluded by the LEP data,
and spontaneous CP violation is not possible in the Higgs sector of the DSTM, at the tree level.
Also, it is found that there is no spontaneous CP violation in the tree level
for $m_{ud}^2 > m_Z^2 \cos \beta \sin \beta$ in Ref. [14].

\subsection{One-loop level}

Now, let us consider the possibility of spontaneous CP violation in the DSTM at the one-loop level.
In some supersymmetric models, the tree-level masses of the neutral Higgs bosons are increased
as much as up to 45 \%, if the radiative corrections are added at the one-loop level.
Thus, the radiative corrections are generally very important in the Higgs sector.
We expect that the Higgs sector of the DSTM would also be significantly affected
by the one-loop contributions.
In our calculation, for simplicity, we take into account only the top quark and scalar top quark loops,
since their contributions are most dominant.

The effective one-loop potential is given as [15]
\begin{equation}
    V_1 = \sum_{l} {n_l {\cal M}_l^4 \over 64 \pi^2}
    \left [ \log {{\cal M}_l^2 \over \Lambda^2} - {3 \over 2} \right ]  \ ,
\end{equation}
where $\Lambda$ is the renormalization scale in the modified minimal subtraction scheme,
${\cal M}_l$ are the field-dependent top and scalar top quark masses,
and $n_l$ are the degrees of freedom arising from color, charge, and spin factors of the particles in the loops,
that is, $n_l = -12$ for top quarks and $n_l =6$ for scalar top quarks.

After the electroweak symmetry breaking,
the top quark mass is given as $m_t = v \sin \beta$ while the stop quark masses are obtained as
\begin{equation}
m_{{\tilde t}_1, \ {\tilde t}_2}^2 =  {1 \over 2} (m_Q^2 + m_T^2) + m_t^2
+ {1 \over 4} m_Z^2 \cos 2 \beta \mp \sqrt{X_t} \ ,
\end{equation}
where
\begin{eqnarray}
X_t & = & \left ( {1 \over 2} (m_Q^2 - m_T^2) + \left ({2 \over 3} m_W^2 - {5 \over 12} m_Z^2 \right ) \cos 2 \beta \right)^2 \cr
& &\mbox{} + m_t^2 \left ( A_t^2 + \mu^2 \cot^2 \beta - 2 \mu A_t \cot \beta \cos \varphi \right ) \ .
\end{eqnarray}
Note the presence of the CP phase $\varphi$ in the scalar top quark masses.

The minimum condition for the vacuum stability with respect to the CP phase $\varphi$ yields
\begin{equation}
m_{ud}^2 = 2 v^2 \left (\epsilon_1 + \epsilon_2 \sin 2 \beta \cos \varphi \right )
+ {3 m_t^2 A_t \mu  \over 16 \pi^2 v^2 \sin^2 \beta} f (m_{{\tilde t}_1}^2,  \ m_{{\tilde t}_2}^2)  \   ,
\end{equation}
where $f (m_{{\tilde t}_1}^2,  \ m_{{\tilde t}_2}^2)$ is defined as
\begin{equation}
 f(m_x^2, \ m_y^2) = {1 \over (m_y^2 - m_x^2)} \left[  m_x^2 \log {m_x^2 \over \Lambda^2} - m_y^2
\log {m_y^2 \over \Lambda^2} \right] + 1 \ .
\end{equation}

Let us express the $3 \times 3$ symmetric mass matrix for the neutral Higgs bosons at the one-loop level as
\begin{equation}
    M = M^0 + M^1  \   ,
\end{equation}
where $M^0$ is obtained from $V_0$ and $M^1$ is obtained from $V_1$.
We note that $M^0$ in this expression is different from $M^0$ obtained previously for the tree-level
masses.
At the tree level, $M^0$ is calculated after the minimum condition with respect to the CP phase is applied.
Therefore, $M^0$ at the tree level does not contain the soft SUSY mass $m_{ud}$.
On the other hand, at the one-loop level, we should calculate  $M^0$ without applying
the minimum condition with respect to the CP phase.
Thus, $M^0$ at the one-loop level would contain $m_{ud}$, which will eventually be eliminated
when we apply the minimum condition with respect the CP phase.

Explicitly, at the one-loop level, the elements of $M^0$ are obtained as
\begin{eqnarray}
M_{11}^0 & = & m_Z^2 \cos^2 \beta + m_{ud}^2 \tan \beta \cos \varphi
+ 2 \epsilon_1 v^2 ( 2 \cos 2 \beta + 1 ) \tan \beta \cos \varphi      \ , \cr
M_{22}^0 & = & m_Z^2 \sin^2 \beta + m_{ud}^2 \cot \beta \cos \varphi
- 2 \epsilon_2 v^2 ( 2 \cos 2 \beta - 1 ) \cot \beta \cos \varphi   \ , \cr
M_{33}^0 & = & {m_{ud}^2 \cos \varphi \over \cos \beta \sin \beta}
- {2 \epsilon_1 v^2 \cos \varphi \over \cos \beta \sin \beta} - 4 \epsilon_2 v^2 \cos 2 \varphi  \  ,   \cr
M_{12}^0 & = &\mbox{} - m_Z^2 \cos \beta \sin \beta - m_{ud}^2 \cos \varphi + 4 \epsilon_1 v^2 \cos \varphi
+ 2 \epsilon_2 v^2 \sin 2 \beta \cos 2 \varphi   \ , \cr
M_{13}^0 & = & m_{ud}^2 \cos \beta \sin \varphi - 6 \epsilon_1 v^2 \cos \beta \sin \varphi
- \epsilon_2 v^2 ( \cos 2 \beta + 3 ) \sin \beta \sin 2 \varphi   \ , \cr
M_{23}^0 & = & m_{ud}^2 \sin \beta \sin \varphi - 6 \epsilon_1 v^2 \sin \beta \sin \varphi
+ \epsilon_2 v^2 ( \cos 2 \beta - 3 ) \cos \beta \sin 2 \varphi   \ .
\end{eqnarray}

Likewise, the elements of $M^1$ are obtained as
\begin{eqnarray}
M^1_{ij} & = & {3 W_i W_j \over 32 \pi^2 v^2}
{g(m_{{\tilde t}_1}^2, m_{{\tilde t}_2}^2) \over (m_{{\tilde t}_2}^2 - m_{{\tilde t}_1}^2)^2}
+ {3 A_i A_j \over 32 \pi^2 v^2}
\log \left ( {m_{{\tilde t}_1}^2 m_{{\tilde t}_2}^2 \over \Lambda^4 } \right ) \cr
& &\mbox{} + {3 \over 32 \pi^2 v^2} (W_i A_j + A_i W_j)
{ \log ( m_{{\tilde t}_2}^2/ m_{{\tilde t}_1}^2)  \over (m_{{\tilde t}_2}^2 - m_{{\tilde t}_1}^2)} + D_{ij}  \  ,
\end{eqnarray}
where
\begin{equation}
 g(m_x^2,m_y^2) = {m_y^2 + m_x^2 \over m_x^2 - m_y^2} \log {m_y^2 \over m_x^2} + 2 \ ,
\end{equation}

\renewcommand\thefigure{FIG. 2a}
\begin{figure}[t]
\begin{center}
\includegraphics[scale=0.5]{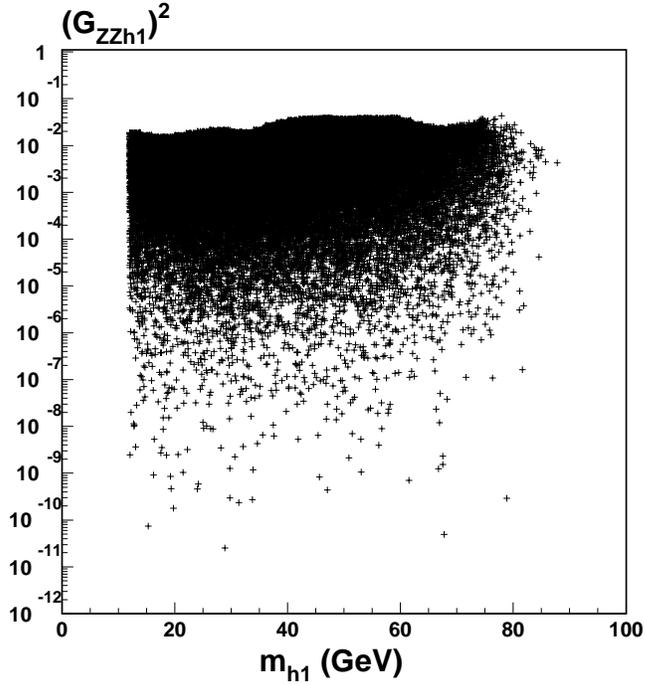}
\caption[plot]{The distribution of $(m_{h_1} , g_{ZZh_1}^2)$,
for each of 48914 points in the parameter region, defined as
$|\varphi| < \pi/2$, $0 < \epsilon_1 < 0.05$, $0 < \epsilon_2 < 0.05$,
$2 < \tan \beta < 30$, $|\mu| < 1000$ GeV, $0 < A_t {\rm ~(GeV)~} < 2000$,
$100 < m_Q {\rm ~(GeV)~} < 1000$, and $100 < m_T {\rm ~(GeV)~} < 1000$.
The marks are all consistent with the LEP data.}
\end{center}
\end{figure}

\begin{eqnarray}
D_{33} & = &\mbox{} - {3 \over 16 \pi^2 v^2}
\left( {m_t^2 \mu A_t \cos \varphi \over \sin^3 \beta \cos \beta} \right)
f(m_{{\tilde t}_1}^2, \ m_{{\tilde t}_2}^2) \ , \cr
& & \cr
D_{11} & = & \sin^2 \beta D_{33} - {3 \cos^2 \beta \over 16 \pi^2 v^2}
\left( {4 m_W^2 \over 3} - {5 m_Z^2 \over 6} \right)^2
f(m_{{\tilde t}_1}^2, \ m_{{\tilde t}_2}^2) \ , \cr
& & \cr
D_{22}
& = & \cos^2 \beta D_{33} - {3 \sin^2 \beta \over 16 \pi^2 v^2}
\left( {4 m_W^2 \over 3} - {5 m_Z^2 \over 6} \right)^2
f(m_{{\tilde t}_1}^2, \ m_{{\tilde t}_2}^2)     \cr
& & \cr
& &\mbox{} - {3 m_t^4 \over 4 \pi^2 v^2 \sin^2 \beta} \log \left ({m_t^2 \over \Lambda^2} \right )  \ , \cr
& & \cr
D_{12}
& = &\mbox{} - \cos \beta \sin \beta D_{33}
+ {3 \cos \beta \sin \beta \over 16 \pi^2 v^2} \left ({4 m_W^2 \over 3} - {5 m_Z^2 \over 6} \right)^2
f(m_{{\tilde t}_1}^2, \ m_{{\tilde t}_2}^2)    \ , \cr
& & \cr
D_{13} & = &\mbox{} - {3 \over 16 \pi^2 v^2}
\left ( {m_t^2 \mu A_t \sin \varphi \over \tan \beta \sin \beta} \right )
f(m_{{\tilde t}_1}^2, \ m_{{\tilde t}_2}^2)  \ , \cr
& &  \cr
D_{23} & = &\mbox{} \tan \beta D_{13}  \ ,
\end{eqnarray}
and
\begin{eqnarray}
A_1 & = & {1 \over 2} m_Z^2 \cos \beta \ , \cr
A_2 & = & {2 m_t^2 \over \sin \beta} - {1 \over 2} m_Z^2 \sin \beta  \ , \cr
A_3 & = & 0   \ ,  \cr
W_1 & = & {2 m_t^2 \mu \Delta_{{\tilde t}_1} \over \sin \beta } + \cos \beta \Delta_{\tilde t}  \ , \cr
W_2 & = &\mbox{} - {2 m_t^2 A_t \Delta_{{\tilde t}_2} \over \sin \beta } - \sin \beta \Delta_{\tilde t}   \ , \cr
W_3 & = & {2 m_t^2 \mu A_t \sin \varphi \over \sin^2 \beta}  \ ,
\end{eqnarray}
with
\begin{eqnarray}
\Delta_{{\tilde t}_1} & = & \mu \cot \beta - A_t \cos \varphi \ , \cr
\Delta_{{\tilde t}_2} & = & \mu \cot \beta \cos \varphi - A_t \ , \cr
\Delta_{\tilde t} & = & \bigg ({4 \over 3} m_W^2 - {5 \over 6} m_Z^2 \bigg )
\bigg (m_Q^2 - m_T^2 + \bigg ({4 \over 3} m_W^2 - {5 \over 6} m_Z^2 \bigg) \cos 2 \beta \bigg)   \  .
\end{eqnarray}
Note that, at the one-loop level, too, non-zero $\varphi$ triggers spontaneous CP violation.

\renewcommand\thefigure{FIG. 2b}
\begin{figure}[t]
\begin{center}
\includegraphics[scale=0.5]{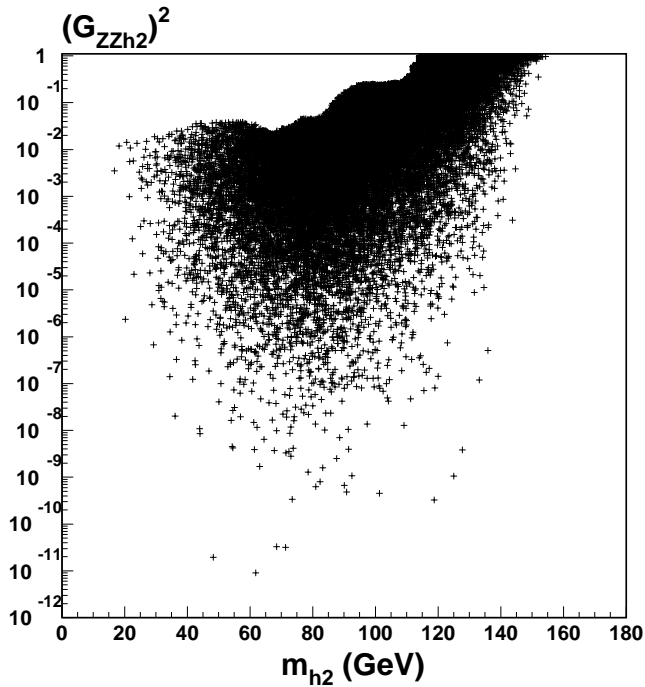}
\caption[plot]{The distribution of $(m_{h_2}, g_{ZZh_2}^2)$. Other captions are the same as Fig. 2a.}
\end{center}
\end{figure}

Now, we investigate whether spontaneous CP violation scenario at the one-loop level is consistent with the LEP data.
For the numerical analysis, we set up a parameter region,
defined as $|\varphi| < \pi/2$, $0 < \epsilon_1 < 0.05$, $0 < \epsilon_2 < 0.05$,
$2 < \tan \beta < 30$, $|\mu| < 1000$ GeV, $0 < A_t {\rm ~(GeV)~} < 2000$,
$100 < m_Q {\rm ~(GeV)~} < 1000$, and $100 < m_T {\rm ~(GeV)~} < 1000$.
Then, we scan $2 \times 10^5$ points in this parameter region, by using the Monte Carlo method.

For each point, we undergo the same procedure as in the tree-level case.
We calculate $m_{h_i}$ and $g_{ZZh_i}^2$ ($i = 1,2,3$), where $g_{ZZh_i}^2$ ($i = 1,2,3$) is
the normalized Higgs coupling coefficient to a pair of $Z$ bosons, normalized by the corresponding SM Higgs coupling coefficient.
The results are shown in Figs. 2a and 2b, where $(m_{h_1} , g_{ZZh_1}^2)$ and $(m_{h_2} , g_{ZZh_2}^2)$ are plotted,
respectively.

The main difference of Figs. 2a and 2b from Fig. 1 is that the constraint from the LEP data is already taken into account in
Figs. 2a and 2b.
The result is shown in Figs. 2a and 2b, where a distribution of 48914 points is displayed.
Thus, those marks above the solid curve of the model-independent upper bound on $g_{ZZH}^2$ are already deleted from
the figures.
Therefore, all of the marks shown in the figures are consistent with the LEP data.
Our analysis shows that the DSTM allows spontaneous CP violation at the one-loop level, for a reasonably wide parameter space.

We find from Fig. 2a that the upper bound on the mass of the lightest neutral Higgs boson of the DSTM
at the one-loop level is about 87 GeV.
From Fig. 2b, we see that $m_{h_2}$ (GeV) at the one-loop level is between 16 and 155.
The mass of the heaviest neutral Higgs boson at the one-loop level is estimated to be
$114 \le m_{h_3}{\rm ~(GeV)~} \le 278$, although the result for $(m_{h_3} , g_{ZZh_3}^2)$ is not shown.

According to the benchmark scenario of maximal CP violation in the MSSM (the CPX scenario),
the LEP2 data might not be suitable to observe the existence of a light Higgs boson
with $m_{h_1} \sim 50$ GeV in the MSSM with explicit CP violation at the one-loop level,
since the strength of the $h_1$ coupling to a pair of $Z$ bosons can be very small [16].
At the LHC, the possibility of observing the light Higgs boson in the MSSM with explicit CP violation
might be investigated via such processes as the gluon fusion process, the vector boson fusion process,
the Higgs-strahlung process, and the $t{\bar t}$-associated Higgs production.
Although the most of the parameter space of the CPX scenario may be investigated for the light Higgs boson via these
processes, there is a narrow window where the $h_1$ investigation is not enough because the coupling coefficients of $h_1 ZZ$, $h_1 WW$,
and $h_1 t{\bar t}$ may be suppressed [17].
Even if the heavier Higgs boson in the MSSM may have large coupling coefficients for $ZZ$, $WW$, and $t{\bar t}$ pairs,
its decay would be very different from the $h_1$ decay.
Thus, for $h_1$, the associated production with a pair of the lighter scalar top quarks might be
the dominant production mechanism, since the trilinear parameter $A_t$ may be as large as 1 TeV
in the CPX scenario [18,19].
A similar analysis may be performed for the Higgs boson in the DSTM with spontaneous CP violation at the LHC.

\section{Conclusions}

We have studied the Higgs sector of the DSTM, which is a two Higgs doublet supersymmetric model with two
effective dimension-five operators, obtained from a power series of $1/M$ in effective action analysis
where $M$ is the energy scale for the new physics beyond the MSSM.
The particle content of the DSTM is similar to the MSSM, but the Higgs phenomenology is distinctively
different.
The MSSM does not accept spontaneous CP violation scenario either at the tree level or at the one-loop level.
On the contrary, we find that the DSTM allows spontaneous CP violation at the one-loop level.

Our calculation is straightforward.
We introduce a physical complex phase as the relative phase between two complex vacuum expectation values,
assuming that the CP symmetry is spontaneously broken.
We calculate the masses of the three neutral Higgs bosons
of the DSTM at the one-loop level, for a reasonably established parameter region.
The masses in GeV are obtained as $10 \le m_{h_1} \le 87$, $16 \le m_{h_2} \le 155$, and $114 \le m_{h_3} \le 278$.
Then, we examine if these results are consistent with the latest experimental limit of
the lightest Higgs boson mass from the LEP data.
The constraint by the LEP data is interpreted by considering $(g^{\rm max}_{ZZ h_i})^2$,
the model-independent upper bound on the Higgs coupling coefficient to a pair of $Z$ bosons,
as a function of the Higgs mass.
In this way, we find that our result for the masses of the three neutral Higgs bosons
of the DSTM at the one-loop level within spontaneous CP violation scenario is allowed by the
LEP data, for a wide choice of parameter values in the established parameter region.

However, we also find that the DSTM cannot accommodate spontaneous CP violation scenario at the tree
level.
Most choice of the parameter values in the parameter region yield unsatisfactory results for
the masses of the three neutral Higgs bosons and their corresponding coupling coefficient to
a pair of $Z$ bosons, unacceptable by the LEP data.
In reality, as we scan $5 \times 10^6$ points in the parameter region by the Monte Carlo method,
no point is consistent with the LEP constraint.

In conclusion, spontaneous CP violation may take place in the DSTM at the one-loop level, but not
at the tree-level, for a reasonable parameter region.
We would like note that the spontaneous CP violation in the DSTM is not radiative CP mixings
because there is a non-trivial CP phase at the tree level.
A similar analysis is under way for explicit CP violation in the Higgs sector of the DSTM [20].

\section*{Acknowledgments}
S. W. Ham thanks P. Ko for the hospitality at KIAS where a part of this work has been performed.
He thanks Kihyeon Cho at KISTI for the collaboration.


\end{document}